\documentstyle[aps,twocolumn,psfig]{revtex}

\begin{document}
\twocolumn[\hsize\textwidth\columnwidth\hsize\csname @twocolumnfalse\endcsname

\title{Role of Disorder on the Quantum Critical Point of a Model for
  Heavy Fermions}
\author{\textsc{{T. G. Rappoport, A. Saguia, B. Boechat and M. A. Continentino}}\\
  Departamento de F\'{\i}sica - Universidade Federal Fluminense\\
  24210-130, Niter\'oi , RJ - Brazil; e-mail: amen@if.uff.br \\}
\date{\today} \maketitle
\begin{abstract}
  A zero temperature real space renormalization group (RG) approach is
  used to investigate the role of disorder near the quantum critical
  point (QCP) of a Kondo necklace (XY-KN) model. In the pure case this
  approach yields $J_{c}=0$ implying that any coupling $J \not = 0 $
  between the local moments and the conduction electrons leads to a
  non-magnetic phase. We also consider an anisotropic version of the
  model ($X-KN$), for which there is a quantum phase transition at a
  finite value of the ratio between the coupling and the bandwidth,
  $(J/W)$. Disorder is introduced either in the on-site interactions
  or in the hopping terms. We find that in both cases randomness is
  irrelevant in the $X-KN$ model, i.e., the disorder induced
  magnetic-non-magnetic quantum phase transition is controlled by the
  same exponents of the pure case. Finally, we show the fixed point
  distributions $P_{J}(J/W)$ at the atractors of the disordered,
  non-magnetic phases.

\end{abstract}

]


Heavy-fermion materials exhibit a variety of interesting phenomena and
different ground states including antiferromagnetic, superconductor,
Kondo-insulator and metallic~\cite{thompson,Aeppli}. Most of the
properties of these systems can be attributed to their proximity to a
magnetic quantum critical point ($QCP$) \cite{livro}. An important and
actual problem in the study of these materials is to understand the
role of disorder on the quantum critical point. This is also a general
problem in statistical mechanics which had a lot of progress in the
last years mainly due to the study of random quantum spin chains
\cite{qc}. For heavy fermions it is particularly relevant to
understand the role of disorder since a standard way of bringing these
systems to the $QCP$ is by doping, which necessarily perturbs the
system \cite{aronson}.

The disordered Kondo model \cite{miranda} and the Griffiths phase
approach \cite{castroneto} have been proposed to deal with disorder in
heavy fermions.  In spite of bringing progress to this field they are,
at this point, mostly phenomenological theories. It is important that
results can be obtained starting from a microscopic Hamiltonian which
is closest to describe heavy fermions and that can incorporate
disorder at a basic level. In this Communication we present a
non-perturbative real space renormalization group approach to the
one-dimensional Kondo necklace ($KN$) model at zero temperature, both
with and without disorder. This model was proposed by Doniach
\cite{Doni} to study heavy fermions and emphasizes magnetic degrees of
freedom~\cite{Fye,Tsu,Julli,Yu} neglecting charge fluctuations. It
incorporates the essential physics of these systems, namely the
competition between Kondo effect and magnetic ordering. The quantum
phase transition of this model is a magnetic transition which is the
actual case in most heavy fermions materials. As far as we know the
role of disorder in the $KN$ model has never been considered and the
present Communication represents a first step towards describing its effects
in heavy fermions starting at a microscopic level. The $KN$ model is
given by the Hamiltonian,
\begin{equation}
H=\sum_{i=1}^{L-1}W_{i}({{\sigma^{x}}}_{i}{{\sigma^{x}}}_{i+1}+{{\sigma^{y}}%
}_{i}{{\sigma^{y}}}_{i+1})+\sum_{i=1}^{L-1}J_{i}{\vec{S}}_{i}.{\vec{\sigma}%
}_{i},\label{hamil}%
\end{equation}
where $\sigma^{\mu}$ and $S^{\mu}$, $\mu=x,y,z$ are spin-1/2 Pauli
matrices denoting the spin of the conduction electrons and those of
the local moments, respectively. The sites $i$ and $i+1$ are
nearest-neighbors on a chain of $L$ sites and $W_{i}$ is an
antiferromagnetic coupling which represents the hopping of the
conduction electrons between neighboring sites. In the present study,
either the on-site interactions between the conduction electrons and
the local moments, $J_{i}>0$ or the hopping $W_{i}$ are uncorrelated
quenched random variables with initial probability distributions,
$P_{J}(J_{i}/W)$ and $P_{W}(J/W_{i})$ respectively. A closely related
model, which we also study here is the $X$-Kondo necklace ($X-KN$)
model where the band of conduction
electrons is represented just by the Ising term, $\sum_{i=1}^{L-1}%
W_{i}{{\sigma^{x}}}_{i}{{\sigma^{x}}}_{i+1}$.

In the pure case our zero temperature RG calculations yield distinct results
for these models. For the $X-KN$ we find an unstable fixed point at a finite
value of $(J/W)$ separating an antiferromagnetic phase from a spin
compensated, Kondo-like phase, which can be fully characterized. For the
$XY-KN$, any interaction $J$ gives rise to a dense Kondo state. The latter
result, $J_{c}=0$, for the $XY-KN$ is in agreement with the most recent
density-matrix renormalization group \cite{caron} and quantum Monte Carlo
calculations \cite{scalettar85}.

The introduction of disorder in the on-site interactions $J$ or in the hopping
terms leads us to consider the renormalization of the probability
distributions, $P_{J}(J/W)$ and $P_{W}(J/W)$. For the $X-KN$ model, disorder
either in the interactions or in the hopping turns out to be an
\emph{irrelevant perturbation at the pure system fixed point}, i.e., the zero
temperature exponents associated with the transition from the
antiferromagnetic ($AF$) phase to the non-magnetic phase in the presence of
disorder are those of the pure case. This is in strong contrast with the Ising
model in a transverse field, which has been used as a model for heavy fermions
\cite{castroneto}, for which disorder is a relevant perturbation at $d=1$
\cite{young} and probably also for $d<4$ \cite{fisher}. The atractor of the
antiferromagnetic phase in the presence of disorder is the same of the pure
case. There is a new atractor associated with the disordered non-magnetic
phase which is characterized by the fixed point probability distributions
$P_{J}^{\ast}(J/W)$ and $P_{W}^{\ast}(J/W)$ at this atractor in each case. In
the $XY-KN$ model the flows are always towards non-magnetic phases, which are
different in the pure and disordered cases. The fixed point distributions
$P_{J}^{\ast}(J/W)$ in the non-magnetic phases of the $X-KN$ and the $XY-KN$
in the presence of disorder are extremely wide. They show huge values of
$(J/W)$ and negligeable weight for small values of the ratio $(J/W)$ which are
progressively being eliminated.

\begin{figure}[!htb]
{\centerline {\psfig{figure=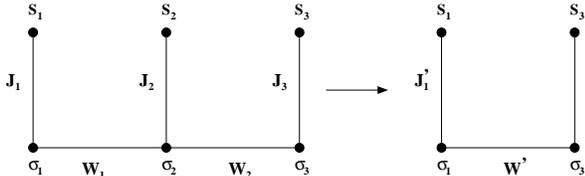,width=8cm}}
\caption{The cells used in the renormalization group transformation and
which correspond to a length scale factor $b=2$.}}
\label{fig1}
\end{figure}
In the context of the real-space renormalization group, the linear chain may
be viewed as $b$ bonds \emph{in series}; $b$ is the scaling factor ($b=2$ in
Fig.~1). For the Hamiltonian given in Eq.~\ref{hamil}, the non-commutation
aspects are present at the cluster level, in the sense that individual spins
are not in a definite state. This can be dealt with by referring to the
density matrix in the basis $|m_{1}m_{2}...m_{n}m_{1}^{\prime}m_{2}^{\prime
}...m_{n}^{\prime}\rangle$, where $S^{x}|m_{i}^{\prime}\rangle=m_{i}^{\prime}%
|m_{i}^{\prime}\rangle$ and $\sigma^{x}|m_{i}\rangle=m_{i}|m_{i}\rangle$, and defining the
renormalization-group (RG) transformation by the mapping of diagonal elements
only. At zero temperature the density matrix is essentially the ground state
projector. This approach has been successfully applied to the transverse Ising
model~\cite{raimundo} and to quantum spin glasses~\cite{bia1}. The disorder is
incorporated through a statistical renormalization group (SRG) treatment where
one follows the effect of a RG transformation on the probability distributions
of the relevant parameters, instead of forcing them to a particular form~
\cite{bia1}. Let us consider the case of random interactions $J$. For a given
configuration of on-site interactions ($\left\{  J_{i}/W \right\}  $) the RG
transformation for the $d=1$ system (Fig.~\ref{fig1}) is defined by,

\begin{equation}
\langle m_{1}m_{3}|\tilde{\rho}^{\prime}(K^{\prime})|m_{1}m_{3}\rangle~=~\langle m_{1}m_{3}
|\tilde{\rho}(K_{1},K_{2},K_{3})|m_{1}m_{3}\rangle\label{eqrho}%
\end{equation}
where $K^{\prime}=(J/W)^{\prime}$ is the renormalized ratio of parameters in
the smallest two-site cell. We have assigned a single index $i$ to each bond
in the original cell $K_{i}={J_{i}/W}$ consistent with an uncorrelated
distribution for these ratios. The matrix elements,
\begin{eqnarray}
\langle m_{1}m_{3}|\tilde{\rho}^{\prime}(K^{\prime})|m_{1}m_{3}\rangle~= \nonumber 
\\~\sum
_{m_{1}^{\prime},m_{3}^{\prime}}\langle m_{1}m_{3}m_{1}^{\prime}m_{3}^{\prime}%
|\tilde{\rho}^{\prime}(K^{\prime})|m_{1}m_{3}m_{1}^{\prime}m_{3}^{\prime
}\rangle\label{eqrhotil1}%
\end{eqnarray}
and
\begin{eqnarray}
\langle m_{1}m_{3}|\tilde{\rho}(K_{1},K_{2},K_{3})|m_{1}m_{3}\rangle~=\label{eqrhotil2}~\\
~\sum_{m_{2},{m^{\prime}}_{1},{m^{\prime}}_{2},{m^{\prime}}_{3}}  
\langle m_{1}m_{2}m_{3}{m^{\prime}}_{1}{m^{\prime}}_{2}{m^{\prime}}_{3}| \nonumber
\\ \rho
(K_{1},K_{2},K_{3})|m_{1}m_{2}m_{3}{m^{\prime}}_{1}{m^{\prime}}_{2}{m^{\prime
}}_{3}\rangle
\end{eqnarray}
are obtained by performing the partial trace on the internal spins keeping
those on the terminal sites (along the chain) fixed (see Fig.~\ref{fig1}). For
the renormalized cell, the left hand side of Eq~(\ref{eqrho}) provides
analytical expressions for the primed variables which are given by,

\begin{equation}
\langle++|\tilde{\rho}^{\prime}(K^{\prime})|++\rangle ~ = ~ -{\frac{1}{4}} \left[  {\frac{
1 - \sqrt{1+16{K^{\prime}}^{2}}}{\sqrt{1+16{K^{\prime}}^{2}}}}\right]
\label{r++}%
\end{equation}
and%

\begin{equation}
\langle-+|\tilde{\rho}^{\prime}(K^{\prime})|+-\rangle ~ = ~ {\frac{1}{4}} \left[  {\frac{
1 + \sqrt{1+16{K^{\prime}}^{2}}}{\sqrt{1+16{K^{\prime}}^{2}}}}\right]
\label{r+-}%
\end{equation}

For the original, larger cell, the right hand side of Eq.~(\ref{eqrho}) is
written in a $64$ by $64$ representation and the calculations have to be
performed numerically. The matching of these quantities above in the two cells
yields in principle two RG equations that should allow to treat $J_{i}$ and
$W_{i}$ as independent random variables. It results however that these
equations reduce to a single one which points out that, as concerns the ground
state of the system, there is a unique relevant variable in the problem,
namely, the ratio $(J/W)$. This is clear in the pure system but turns out to
be the case also in the presence of disorder as is further discussed below

For convenience let us define,%

\begin{equation}
Z_{\pm}=\langle++|\tilde{\rho}|++\rangle\pm\langle-+|\tilde{\rho}|+-\rangle\label{smaismenos}%
\end{equation}
for both $b=1$ and $b=2$ cells. We then match the quantity $Z_{-}^{\prime
}/Z_{+}^{\prime}=Z_{-}/Z_{+}$, which is a scale invariant and finally obtain,
\begin{equation}
K^{\prime}={\frac{1}{4}}\sqrt{\left(  {\frac{Z_{+}}{Z_{-}}}\right)  ^{2}%
-1}\label{kl}%
\end{equation}
which is our basic renormalization group equation. 
We emphasize that although we have considered different scale
invariants to match in the distinct cells we always find the same RG equation
or trivial ones \cite{zvi}. We obtain below that the cases of disorder in the
interactions or in the hopping yield the same phase diagram. These results are
consistent with the existence of a single random variable in the problem, the
ratio $(J/W)$, that suffices to characterize the ground state phase diagram. 
Furthermore, as we take $K_i=J_{i}/W$ or $G_i=J/W_{i}$ as random
variables, $W$ or $J$ respectively, are uniform only at the level of the
cells and not in the chain generated by the RG procedure.

We first consider the pure case without disorder. For the $X-KN$ these
equations yield a quantum critical point at $K_{c}=G_{c}=(J/W)_{c}=0.21$
separating an antiferromagnetic from a non-magnetic, Kondo singlet or
spin-liquid phase. The critical exponents associated with this transition can
be obtained in the usual way. For the correlation length, given by $\xi
\propto|\delta|^{-\nu}$, with $\delta=K-K_{c}$, we obtain the exponent,
$\nu=2.24\approx9/4$. The dynamic exponent $z$ is obtained from the finite
size scaling relation for the gaps for excitations in the two cells, i.e.,
$(\Delta E/\Delta E^{\prime})_{K_{c}}=(L^{\prime}/L)^{z}=b^{z}$, at the
quantum critical point, where $b=2$ is the length scale factor. We get
$z=1.28$. Using the quantum hyperscaling relation $2-\alpha=\nu(d+z)$ we can
also obtain the exponent $\alpha=-3.12$ which is negative but large in module.
The non-magnetic phase, for $K > K_{c}$, is a \emph{spin-liquid} phase
characterized by a gap $\Delta E=|\delta|^{\nu z}$ which vanishes at the QCP
with the gap exponent, $s=\nu z=2.28$. For the $XY-KN$ model we find that any
finite coupling $J$ gives rise to a non-magnetic, Kondo-like phase, i.e.,
$K_{c}=0$.
\begin{figure}[!htb]
{\centerline {\psfig{figure=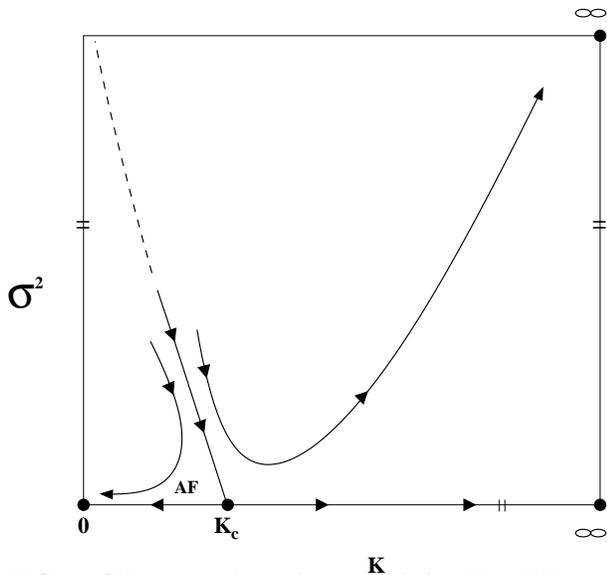,width=8cm}}
\caption{Schematic phase diagram of the $X-KN$ model obtained from our
renormalization group method. The arrows give the flow of the RG equations.
The fixed point at $K_c=(J/W)_c$, $\protect\sigma^2=0$ is that of the pure
system. }}
\label{fig2}
\end{figure}

The effects of disorder in the on-site interactions in the pure Kondo necklace
model, are considered taking, for convenience, a normalized initial
probability distribution, $P(K_{i})=(K_{i}^{\alpha-1}/\beta^{\alpha}%
\Gamma(\alpha))\exp(-K_{i}/\beta)$. This is a single peaked distribution with
an average value $\langle K\rangle=\alpha\beta$ and a variance $\sigma^{2}=\langle(K-\langle K\rangle)^{2}%
\rangle=\alpha\beta^{2}$. Notice that $K\geq0$, i.e., only positive
(antiferromagnetic) values of the on-site interactions are considered. We
start the iteration of the RG equation, choosing three bonds distributed
according to $P(K_{i})$ above, to feed the recursion relations and generate a
new value $K^{\prime}$. This procedure is repeated up to $10^{6}$ times and
yields a new renormalized distribution $P^{\prime}(K^{\prime})$ which will be
used to feedback the recursion relation in the next step of the
renormalization process. The evolution of the distribution is then followed as
the renormalization process goes on. We emphasize that the exact form of the
initial probability distribution is not relevant and the different starting
distributions we used always evolve to the same fixed point form associated
with the atractor of the corresponding phase.

We next consider adding randomness to the on-site interactions of the
pure $X-KN$ model. The zero-temperature RG flow diagram as a function
of $\langle K\rangle=K$ and $\sigma^{2}$ is shown in Fig.~2. In the
presence of disorder the system still exhibits two phases: the Ising
antiferromagnetic phase ($X-AF$) and a disordered, non-magnetic
phase. The atractor of the antiferromagnetic, $X-AF$, phase in the
presence of disorder \emph{is the same} of the pure $AF$ phase.  The
probability distribution in this ordered phase is single peaked and
becomes sharper, i.e., $\sigma^{2}\rightarrow0$ as the mean value
approaches the value $\langle K\rangle=0$ under successive
renormalizations (see Fig.3). On the other hand, it is clear from the
phase diagram in Fig.~2 that the atractor of the non-magnetic phase in
the presence of disorder is different from the atractor of the
non-magnetic, spin-liquid phase of the pure case.


\begin{figure}[!htb]
{\centerline {\psfig{figure=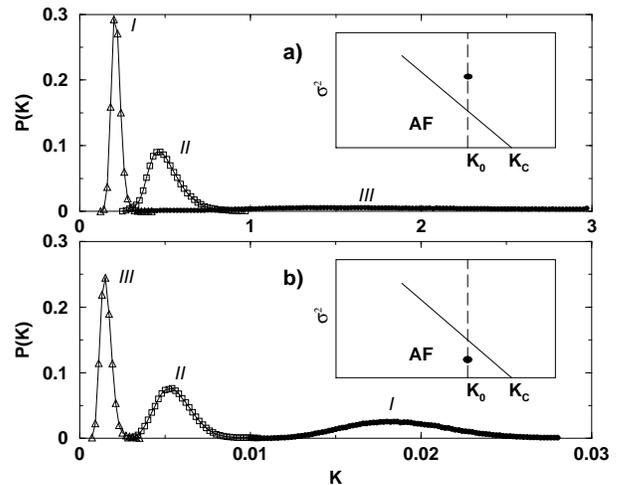,width=8cm}}
\caption{The probability distributions $P_J(J/W)$ at two sucessive steps of
the statistical renormalization procedure ($I$ and $II$ respectively) as the
system iterates to the atractor of the disordered, non-magnetic phase ($a$)
and to the atractor of the antiferromagnetic phase ($b$) of the $X-KN$
model. The insets show the initial points of the RG processes. The critical
boundary in Fig. \ref{fig2} separates the two types of flow shown here.}}
\label{fig3}
\end{figure}

The phase transition to the ordered magnetic phase in the presence of disorder
is governed by the \emph{same} critical exponents of the pure case that we
have calculated before. This is a consequence that the flow along the critical
line is towards the semi-stable fixed point of the pure system at $K_{c}%
=0.21$, $\sigma=0$ which therefore determines the universality class of the
transition, even in the presence of disorder. The irrelevance of disorder
\emph{in the renormalization group sense} is consistent, according to Harris
criterion, with the negative value of the exponent $\alpha=-3.12$ we have
obtained previously. The critical line close to the QCP vanishes as,
$\sigma_{c}^{2}\propto|\delta|=|K-K_{c}|$. The exact form of the critical
frontier in the region represented by the dashed line is not numerically
accessible. Most probably it approaches asymptotically the axis $K=0$ as there
should be no critical value for this type of disorder to destroy the $X-AF$
phase at $K=0$. Notice that disorder is detrimental to antiferromagnetism,
reducing the region of the phase diagram where this can be found.
We have also carried out the statistical RG procedure for the case of disorder
in the hopping terms using the same initial distributions, but now for the
random variable $G_{i}=J/W_{i}$. 
It turns out that the same phase diagram of
Fig.2 is obtained. 

The effects of disorder in the on-site interactions on the
$XY-KN$, which in the pure case has only a non-magnetic Kondo phase, are quite
distinct, from the RG point of view, compared with those in 
the anisotropic $X-KN$ model. 
Firstly, disorder is a relevant
perturbation at the $K_{c}=0$ fixed point since for any initial values of $\langle K\rangle
\not = 0$ and $\sigma^{2}$, the renormalized variance always increases in the
renormalization process. Next, the flow of the distribution $P(J/W)$ towards
the atractor of the random Kondo phase is much faster than for the $X-KN$. The
phase diagram of the $XY-KN$ has only two atractors, that of the pure Kondo
phase at $\sigma=0$, $J/W=\infty$ and that of the random Kondo phase at
$\sigma=\infty$, $\langle K\rangle=\infty$. This is the first time this result has been found
using a real space RG. The same considerations apply for the case 
of off-diagonal disorder, i.e., when the hoppings are random, 
with the replacement $\langle K\rangle \rightarrow \langle G\rangle$.

In the non-magnetic phases, of both the disordered $X-KN$ and $XY-KN$ models,
for the case of disorder in the interactions $J_{i}$, the distributions
$P_{J}(K)$ become extremely wide under iteration as shown in Fig. 3. Small
values of $K$ are being eliminated and huge values for this ratio are being
generated. Note that this behavior of $P_{J}(K)$ does not imply the absence of
low energy excitations since large values of $K$ can occur for $J$ finite and
$W \rightarrow 0$ \cite{Julli}. In fact the only meaningful 
variable in the present approach
is the ratio of these quantities. These disordered phases are better
characterized by the distribution of gaps for excitations which however
can not be obtained in the present approach since it does not take into
account excited states.

In summary, we have studied the effect of disorder in the $KN$ model for heavy
fermion systems. The anisotropic version of the model, the $X-KN$, is
appropriate to describe systems where the ordered magnetic phase has a strong
Ising character. In this case for pure systems we have found a quantum
critical point, at a finite value of $(J/W)$, separating the ordered magnetic
phase from the spin-liquid. Remarkably, we have obtained that at $d=1$, where
the effects of disorder are stronger, this turns out to be an irrelevant
perturbation at the pure system quantum critical point. This holds for both
cases of initial disorder in the interactions $J_{i}$ or in the hoppings $W_{i}$. 
The method used here has the great advantage of being non-perturbative which 
allows to the treat the effects of weak disorder in the neighborhood of the 
quantum critical point of the pure system.
Our results are in agreement with the robustness of
the quantum critical point approach description of non-Fermi liquid behavior
in systems where disorder is clearly present. The present RG method 
characterizes the different phases in terms of their atractors. In particular,
in the disordered phases the same atractor is reached independently of the
point of the phase diagram we start the statistical iteration procedure.
Consequently the present approach does not yield information on the existence
of rare clusters of the incipient magnetically ordered phase which
characterizes the so-called Griffiths phases. This is a feature of the method
which considers only the ground state of the cells and no excited states are
involved in the calculations. Another consequence of this feature is the
existence of a single variable, the ratio $(J/W)$ which suffices to
characterize the phase diagram of the system even in the presence of disorder.
This approach however can be extended to include excited states and to
calculate the distribution of gaps in which case we expect to find evidence of
Griffiths phases \cite{mucio}.

\acknowledgments
The authors are grateful to the Brazilian agencies FAPERJ, CNPq and CAPES for
financial supports.

\end{document}